\documentclass[a4paper, 10pt]{article}
\usepackage{a4}
\usepackage{acfm}
\usepackage[dvips]{graphicx}
\usepackage{indentfirst}
\usepackage{layout} 
\usepackage{subfigure} 
\usepackage{verbatim} 
\usepackage{multicol}
\title{Flux scaling and plume structure in high R\MakeLowercase a - high S\MakeLowercase c turbulent convection} 
\author{Baburaj A. Puthenveettil and \underline{Jaywant H
  Arakeri}} 
\affiliation{Department of Mechanical Engineering, Indian
  Institute of Science,\\ Bangalore, India 560012}
\received{\today} 
\acfmlogo{Proceedings of the Tenth Asian Congress of Fluid Mechanics \\
17--21, May 2004, Peradeniya, Srilanka.}
\begin{document}
\begin{abstract}
  The arrangement of brine above water across a micro porous permeable
  membrane is used to study high Rayleigh Number($10^{11}- 10^{10}$)
  high Schmidt number(650)turbulent convection. The flux shows
  4/3$^{rd}$ scaling with line plume as the near wall coherent
  structures. Shifting of multiple large scale flow cells result in
  changing near membrane mean shear directions for large aspect
  ratios. Lower aspect ratios show single large scale flow cell and
  constant sense of mean shear.

\end{abstract}
\maketitle
\section{Introduction}
In Rayleigh - Benard convection , where buoyancy is the only source of
motion, the non dimensional heat flux, Nusselts number (Nu) can be
expressed as a function of the independent non dimensional parameters
viz. Rayleigh number (Ra) which expresses the ratio of buoyancy forces
to the dissipative effects, Prandtl number (Pr), a fluid property and
Aspect Ratio (AR = Length/Height), a
geometric parameter.  When Rayleigh number is high, turbulent
convection occurs with a well-mixed core and surface boundary layers.
The classical 4/3rd law which states that flux $\sim\,\Delta T ^
{4/3}$ (ie.$ Nu \sim Ra^{1/3})$ is obtained from the assumption that
transport process is determined only by the near wall boundary layers
and hence is independent of layer height h.  At the same time, as very
high Ra implies negligible dissipative effects, dimensional
consistency demands $Nu \sim (Ra Pr)^{1/2}$ ~\cite{cg} .  The
experimentally observed scaling law is $Nu \,=\,K\, Ra^n$ where n is
slightly less than 1/3, resulting in a weak dependence of flux on
layer height.  At $Ra >10^8$ the system is seen to generate a
large-scale mean flow which modifies the near wall diffusive boundary
layers by shear effects, and is expected to be the reason for the
observed value of n .  Plumes play the major role in transporting heat
in this regime from the boundary layer\cite{tjfm}.  Various
phenomenological explanations, each involving major assumptions, viz.
a plume dominated mixing zone\cite{cg}, a turbulent shear boundary
layer\cite{shr}, a Blasius laminar boundary layer with dominant
balance of bulk and boundary layer dissipation\cite{gl} etc seem to be
able to obtain the observed scaling.

The Prandtl number dependence of Nu at high Pr is not clear. The only
studies are that of Goldstein\cite{gcs} using naphthalene mass
transfer technique at Schmidt number of 2750 and of Askenazi and
Steinberg\cite{as} at Pr of 93. The $Ra^{1/3}$ scaling was observed by
Goldstein till Ra $<\, 10^{13}$, while Askenazi's studies seem to
support an exponent less than 1/3. At very high Pr, as the viscous
boundary layers reach their limiting thickness, it is expected that Nu
becomes independent of Pr\cite{gl2}.  In this case the flux is
expected to follow the 4/3 law as the strength of large scale flow
reduces with increasing Prandtl Number.  Hence, in adition to the
large number of unresolved issues about the nature of high Rayleigh
number turbulent free convection, few studies exist for the high
Prandtl Number regime.  Further, very little is known about the nature
of near wall coherent structures in this regime.

We study high Ra turbulent free convection driven by density
difference across a thin permeable horizontal partition separating two
tanks of square plan form cross sections.  The gravitational potential
due to a heavier fluid(brine) above a lighter fluid(water) across the
partition drives the flow, which is resisted by the presence of the
micro porous partition. At low pore sizes in the membrane, the
transport across the partition would become diffusion dominated, while
the transport above and below the partition becomes similar to
turbulent convection above flat horizontal surfaces.  As molecular
diffusivity (D) of NaCl is about 100 times lower than temperature
diffusivity, larger values of Ra and Sc are achieved through this
arrangement for similar driving density potentials.  The structure of
convection in this case can easily be visualised. In this paper we
report the flux scaling and the nature of near wall coherent
structures in high Rayleigh number ($\sim10^9\, \mathrm{to}\,
10^{13}$) high Schmidt number( $\sim 650$) turbulent convection.

\section{Experimental setup and measurements}
\label{sec:exper-proc}
\begin{figure}[tbp]
\parbox{0.5\textwidth}{
  \centering
  \includegraphics[width=0.415\textwidth]{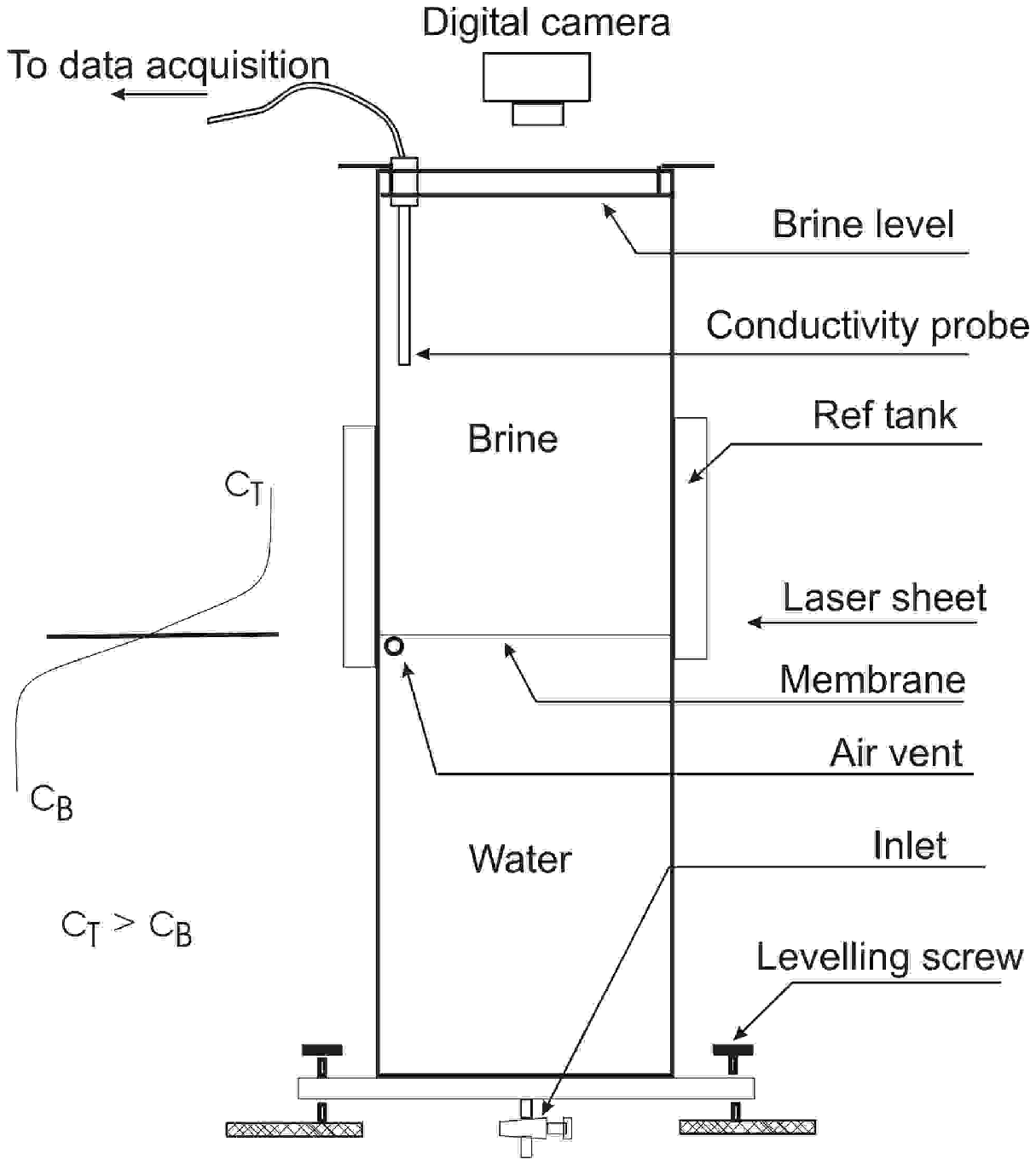}
\caption{Experimental setup}
  \label{fig:setup}}
\parbox{0.5\textwidth}{
  \centering
  \includegraphics[height=0.175\textwidth]{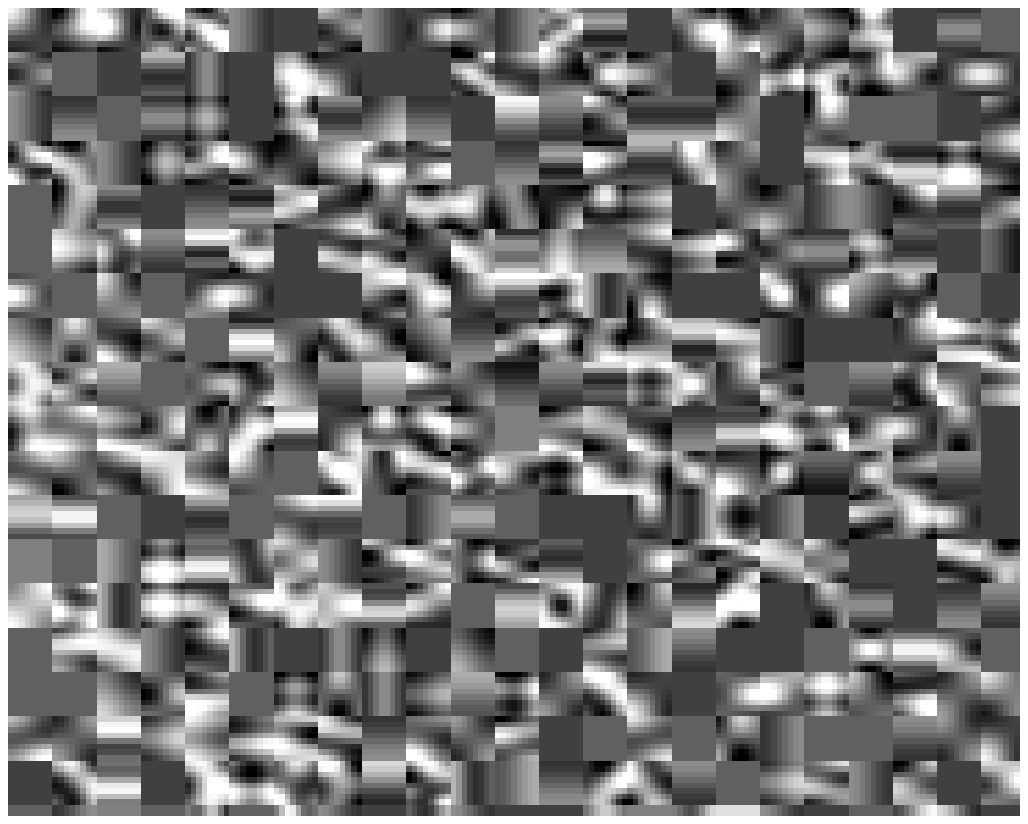}
  \caption{SEM image of Pall Gellman~\texttrademark  membrane}
  \label{fig:pgc}\vspace{0.12cm}
  \includegraphics[width=0.25\textwidth]{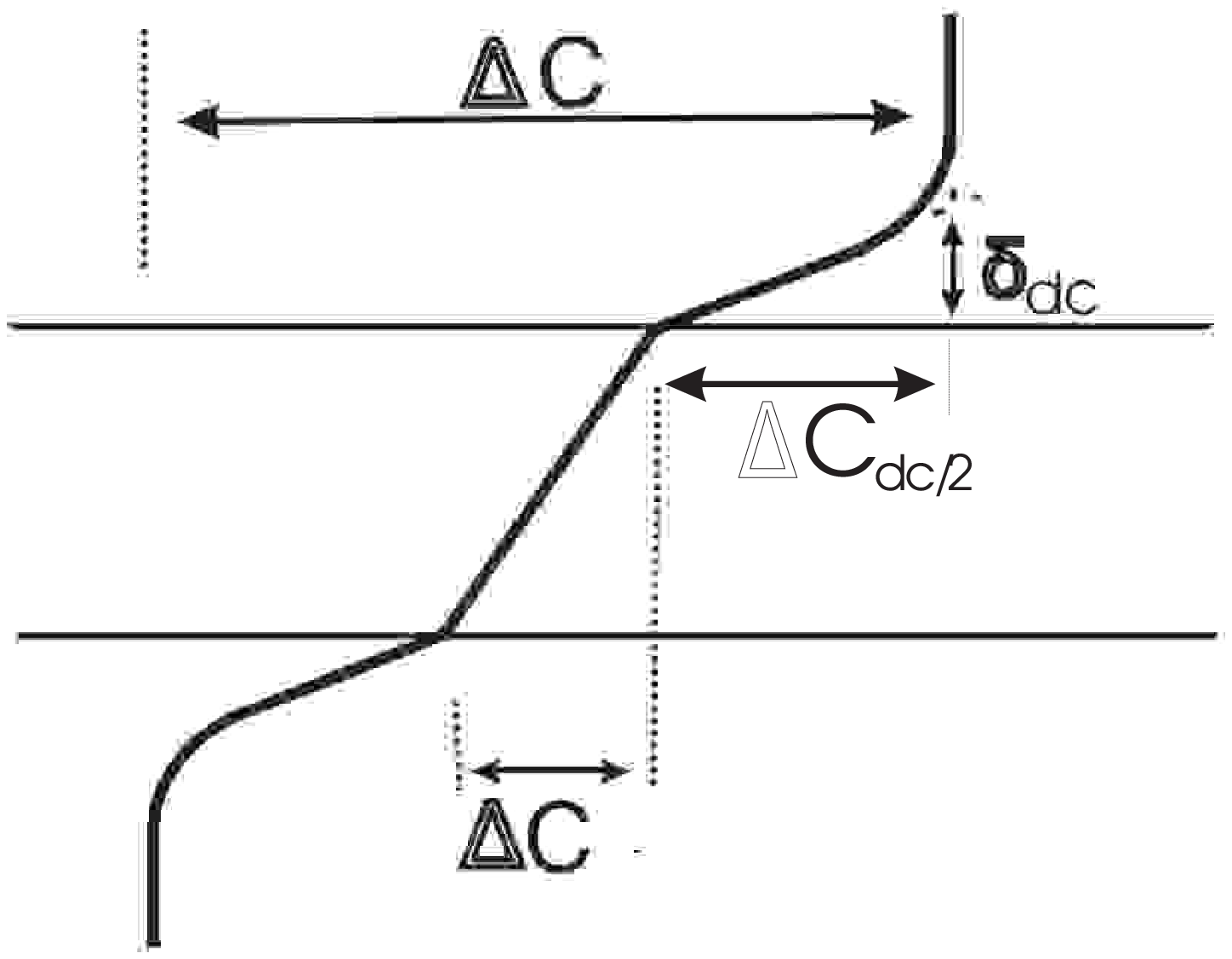}
  \caption{Diffussion drop across the membrane}
  \label{fig:deltadiff}}
\end{figure}
A schematic representation of the experimental setup is shown in
Figure~\ref{fig:setup}. The test section, made of 8mm float glass, has
two vertical compartments,with ground mating edges and a permeable
partition fixed in between them.  Figure~\ref{fig:pgc} shows a
($1000\times$) scanning electrode microscope (SEM) image of Pall
Gelmann~\texttrademark NX29325 membrane disc filters used in the
experiments.  These are bilayer membranes made of nylon66 with
$0.45\mu$ mean pore size and thickness of $142.24 \mu$.  The solidity
of the membrane was calculated as 0.6 from the SEM images by computing
the occupied fraction of pixels of a binary image generated from
figure~\ref{fig:pgc}, based on a suitably chosen threshold.

The top tank is filled with brine after the bottom tank is filled
distilled water. To reduce initial mixing, a temporary tank with
sponge bottom is kept over the membrane while the top tank is filled,
the removal of which initiates the experiment. A thin transparent
plexiglass sheet floating over the brine level prevents evaporation
and produces similar boundary conditions for the two compartments. The
side glass compartments hold distilled water to reduce excessive
refraction of the laser beam during visualisation.  The assembly is
mounted on a levelling table so that the partition can be made
horizontal.  Test section cross sections of 15 cms$\times$ 15cms, and
10cms $\times$ 10cms with height 23 cms, ie aspect ratios of 0.652 and
0.435 were used.

The flux is estimated from the transient measurement of the the top
tank concentration. Assuming that the fluid in both the compartments
is well mixed in the region away from the partition, and using the
mass balance at any time
$C_T(t)\,V_T\,+C_B(t)\,V_B\,=\,C_T^0\,V_T+C_B^0\,V_B$, with $C_B\,^0$
= 0, the concentration difference is
\begin{equation}
  \label{eq:deltac}
  \Delta C(t)\, =\,C_T(t)\,-\,C_B(t)\,=\,(1+\frac{V_T}{V_B})C_T(t)\,-\,C_T^0\,\frac{V_T}{V_B}
\end{equation}
Here, $V_T$ is the top tank solution volume, $V_B$ the bottom tank
solution volume, $C_T(t)$ is the top tank concentration, $C_B(t)$ is
the bottom tank concentration and the superscript $^0$ denotes initial
values.  The nett flux of NaCl across the partition at any instant is
given by the rate of change of top tank concentration as
\begin{equation}
  \label{eq:flux}
  q\,=\,-h\,\frac{d\,C_T}{dt}
\end{equation}
where h is the top compartment height. Hence, $\Delta C$ and flux can
be calculated from the transient measurement of concentration of top
tank fluid. 

The concentration of NaCl in the top tank is estimated from the
measurement of electrolytic conductivity of the top tank fluid.  The
conductivity measurements are made by ORION SENSORLINK~\texttrademark
PCM100 conductivity measurement system~\cite{orm}, with a 2
electrode conductivity cell, model ORION~\texttrademark 011050, with
automatic temperature compensation. The probe was calibrated before
each experiment. As calculation of $\frac{dc }{dt} $ directly from the
measured $C_T$ vs t distribution results in excessive errors due to
data noise an exponential decay fit is used calculate the derivative
$\frac{dC_T}{dt}$.

The plume structure was visualised by Laser Induced fluorescence of
Sodium Flourescein dye. The bottom tank solution was tagged with the
dye and a horizontal laser sheet(Spectra-Physics, Stabilite
2017\texttrademark Ar-Ion,5W ) was passed just above the partition.
The dye in the bottom solution, while convecting upwards, fluoresces
on incidence of the laser beam to make the plume structure visible.
The quantity of dye (1.2ppm) was chosen to give sufficient
fluorescence intensity without affecting the measured conductivity.
The fluorescence images were captured on a digital camera (SONY PCR
9E) after cutting off the laser light wave lengths using a yellow
glass filter, Coherent optics OG-515. 
\section{Results and Discussion}
\label{sec:results}
\subsection{Flux}
\begin{figure}[tbp]
\parbox{0.5\textwidth}{
 \centering 
 \includegraphics[width=0.5\textwidth,height=0.45\textwidth]{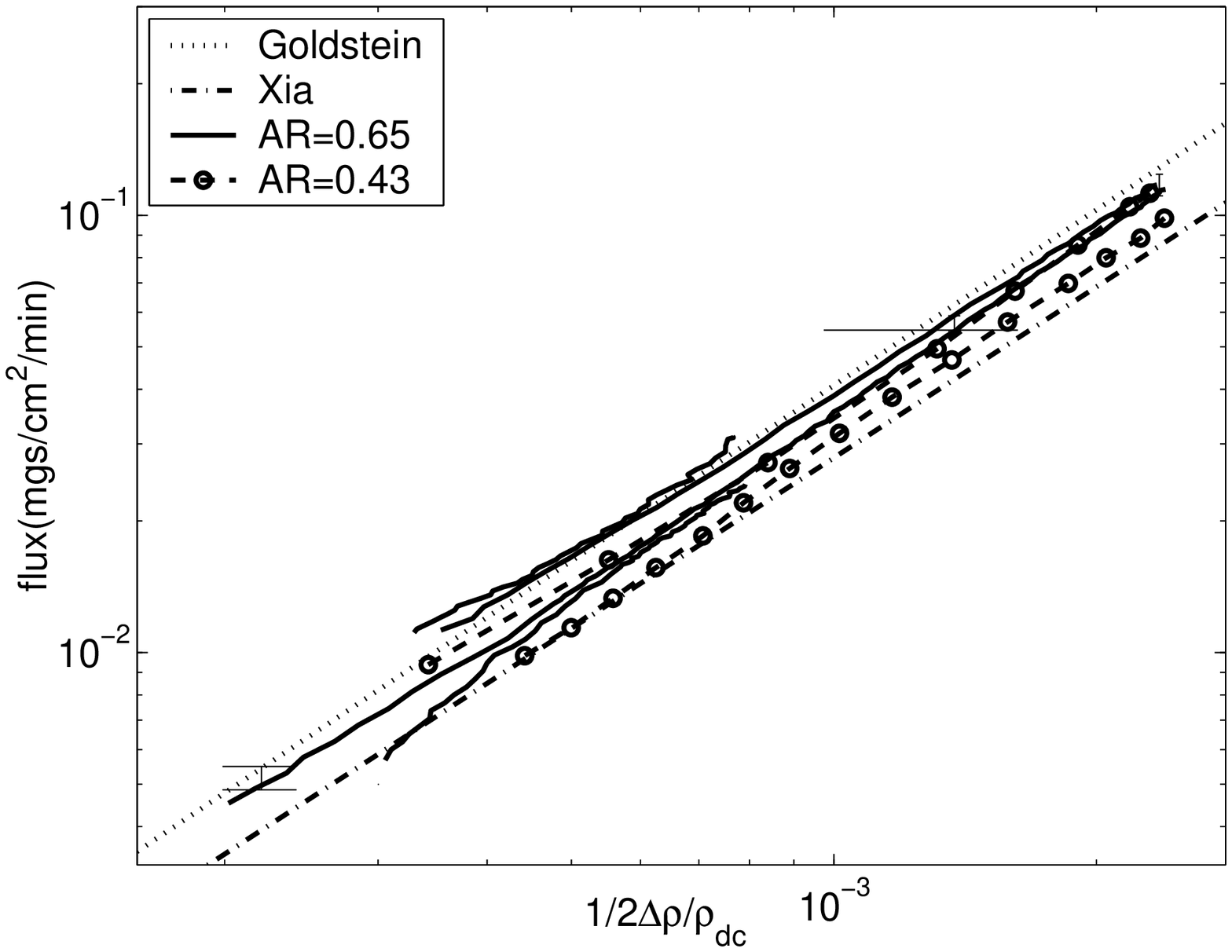}
 \caption{\small Variation of flux with $\frac{1 }{2 } \left(\frac{\Delta\rho}{\rho}\right)_{dc} $ }
\label{fig:flux}}
\parbox{0.5\textwidth}{
\centering
 \includegraphics[width=0.5\textwidth,height=0.45\textwidth]{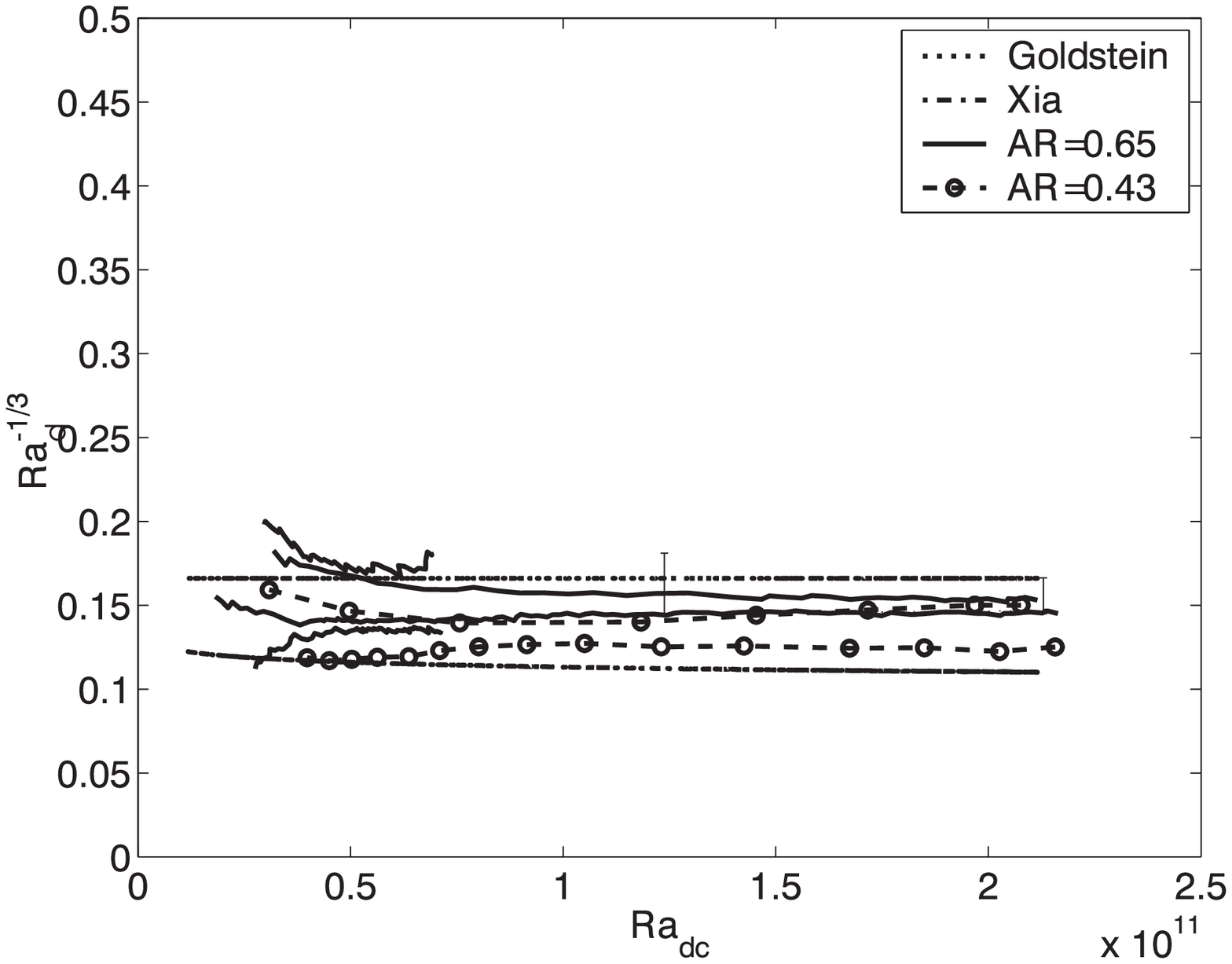}
   \caption{\small Variation of $Ra_{\delta}^{-1/3}$ with$Ra_{dc}$ }
\label{fig:radc}}
\end{figure}

To compare the scaling of flux with that of high Rayleigh number
turbulent free convection, the driving potential in the present case
is corrected for the concentration drop occuring across the partition,
as shown in figure~\ref{fig:deltadiff}. This concentration drop is
$\Delta c_m\,=\,q\,l_m/( \Gamma\,D)$ where $\Gamma$ = open area factor
of the membrane, $l_m$ = the thickness of the membrane.  Therefore,
effective concentration difference on one side of the partition
$\Delta c_{dc}/2\,=\,\left(\Delta c\, -\,\Delta c_m\right) /2$. We
look at the scaling of flux with
$\left(\Delta\rho/\rho\right)_{dc}/2\,=\, \beta\Delta
C_{dc}/2$ in the form of an alternate representation of flux.
Hereafter, the subscript $_{dc}$ denotes this corrected effective
concentration difference.

Figure~\ref{fig:flux} shows the variation of flux with the effective
driving potential obtained in the experiments. The calculations were
conducted after correcting the measured concentration curve for
systematic shifts due to change in probe linearity. The error bars,
which include random and systematic errors, for flux and $\Delta C$
shown in the plot indicate the extent of possible variation. The plot
includes the results for experiments with a starting concentration of
$C_T^0\,=\,10g/l$ at the two aspect ratios of 0.652 and 0.435, as well
as experiments with starting concentration of $C_T^0\,=\,3g/l$ for
AR=0.652. Therefore, within the present accuracy of the measurements,
the flux seems to be independent of the starting concentration and the
aspect ratio. The figure also shows the corresponding results of
Goldstein\cite{gcs} and Xia\cite{xia}. 

Theerthan\cite{tapf} have shown that a more appropriate
representation of non dimensional flux in turbulent free convection is
in terms of $Ra_\delta^{-1/3}$, where $Ra_\delta$ is the Rayleigh
Number based on the diffusion layer thickness, $\delta_{d}$. The
diffusion layer thickness for the current case is calculated as
$\delta_{dc}\,=\,\left(D\Delta c_{dc}/2\right)/q$.
$Ra_{\delta}^{-1/3}$can be written as a ratio of two fluxes as,

\begin{equation}
  \label{eq:radc}
Ra_{\delta}^{-1/3}\,=\,q/\left(\frac{D\Delta C_{dc}/2}{Z_w} \right),\qquad
\mathrm{where,\,} Z_w\,=\,\left(\frac{\nu D }{g\beta \Delta C_{dc}/2}\right)^{1/3} 
\end{equation}
is a near wall length scale for turbulent free convetion\cite{th}.
This representation does not have the layer height as a parameter and
hence is a better representation in turbulent free convection where
near wall phenomena decides the flux.  $Ra_\delta^{-1/3}$ varies only
between 0.1 and 0.3 for a wide range of Ra and various types of free
convection, being a reflection of the fact that heat flux in turbulent
free convection for a given $\Delta T$ and fluid does not vary much.
It could be noticed from (\ref{eq:radc}) that $Ra_\delta^{-1/3}$ is a
constant if flux scales as $\Delta C^{4/3}$.

Figure~\ref{fig:radc} shows the variation of $Ra_{\delta}^{-1/3}$ with
Rayleigh number using the effective concentration difference,
$Ra_{dc}$ for the same experiments as in figure~\ref{fig:flux}.  The
line $Ra_\delta^{-1/3}$ = 0.166 is obtained from the correlation $Nu\,
=\, 0.066 Ra ^{1/3}$ of Goldstein\cite{gcs} at Pr=2750. The relation
of Xia , Nu=$0.14Ra^{0.297}Pr^{-0.03}$ is also plotted in the figure.
The current experimental values of $Ra_{\delta}^{-1/3}$ are nearly
constant implying that the flux scales as the $4/3^{rd}$ power of
$\Delta C_{dc}$.  Goldsteins value of $Ra_\delta^{-1/3}$ falls within
the error range of the current experiments. The deviation from the
constant $Ra_{\delta}^{-1/3}$ for very low $Ra_{dc}$ seen in
figure~\ref{fig:radc} cannot be inferred to be from the change in the
flux scaling, as the errors involved in calculating
$\frac{\Delta\rho}{\rho}$ becomes large when the concentration
differences between the tanks tend to zero.  Therefore, we conclude
from the present experiments that Rayleigh Numbers of $10^{11}
-10^{10}$ at Schmidt number of 650, the flux scales very nearly as
$\Delta C ^{4/3}$, and the effect of Schmidt number on flux is
negligible. To get a better understanding of the phenomena, we now
study the near wall dynamics of coherent structures in the current
experiments.
\subsection{Coherent structures}
\label{sec:coherent-structures}
\begin{figure}[!tbp]
\parbox[t]{0.45\textwidth}{
\includegraphics[width=0.45\textwidth]{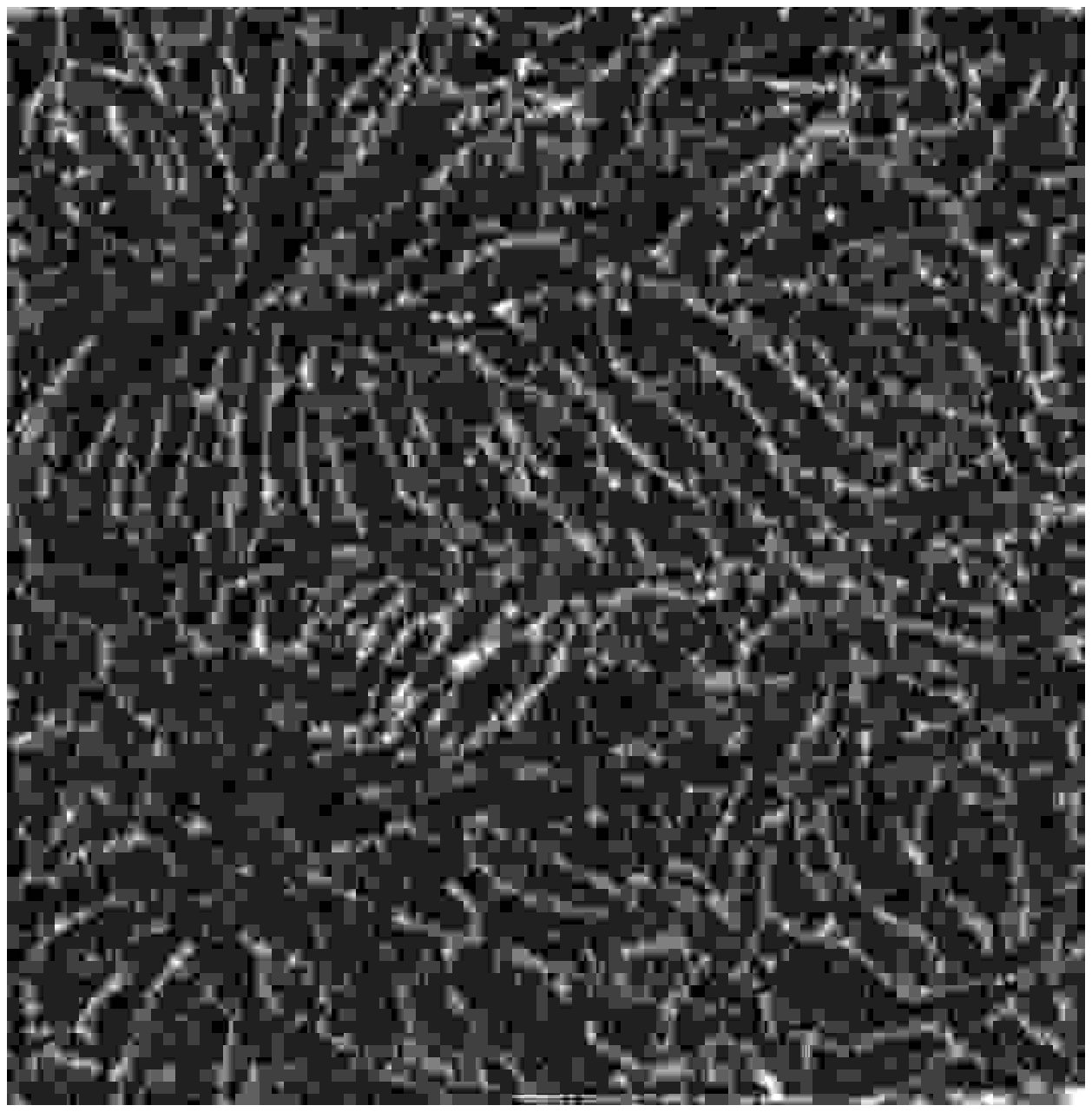} 
\includegraphics[width=0.45\textwidth,height=0.2010\textwidth]{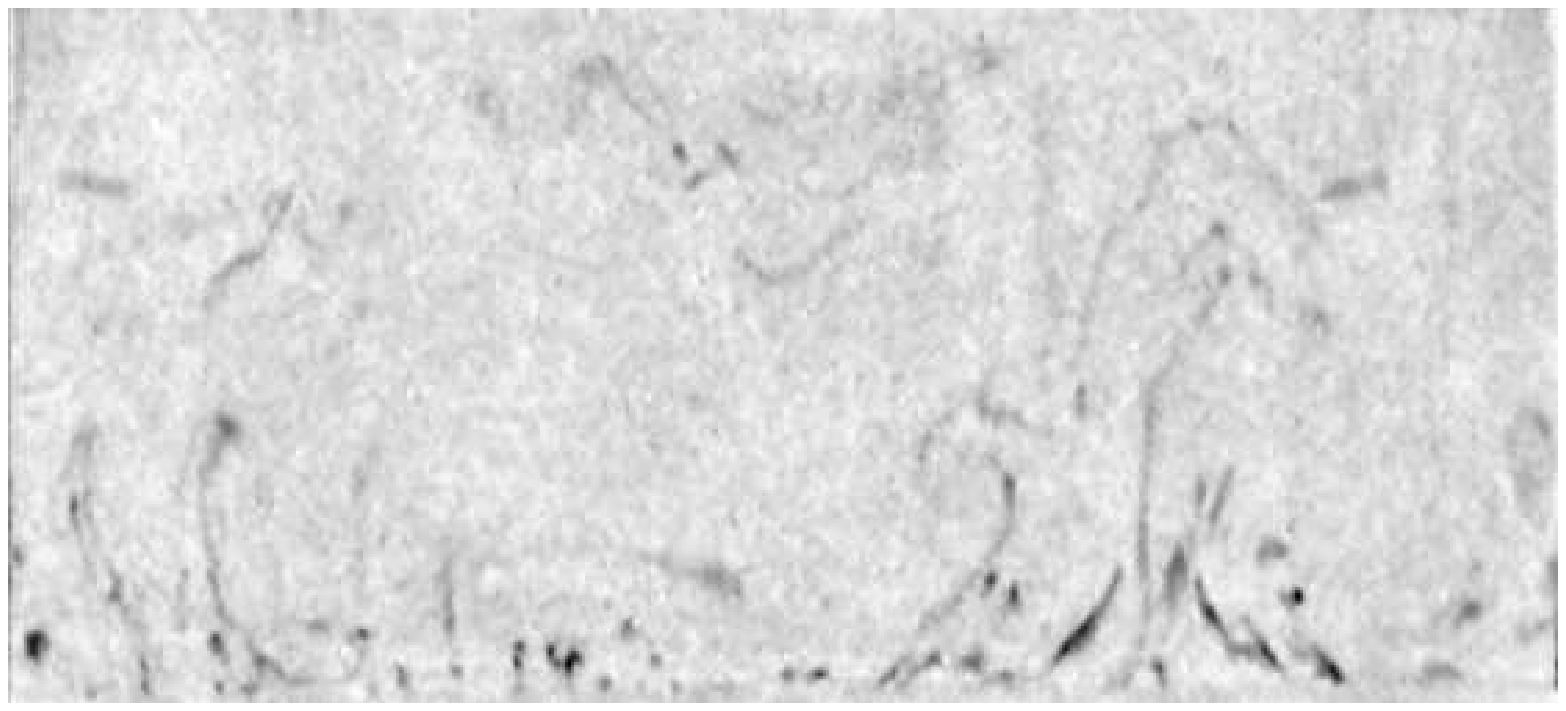} 
\includegraphics[width=0.55\textwidth,height=0.225\textwidth]{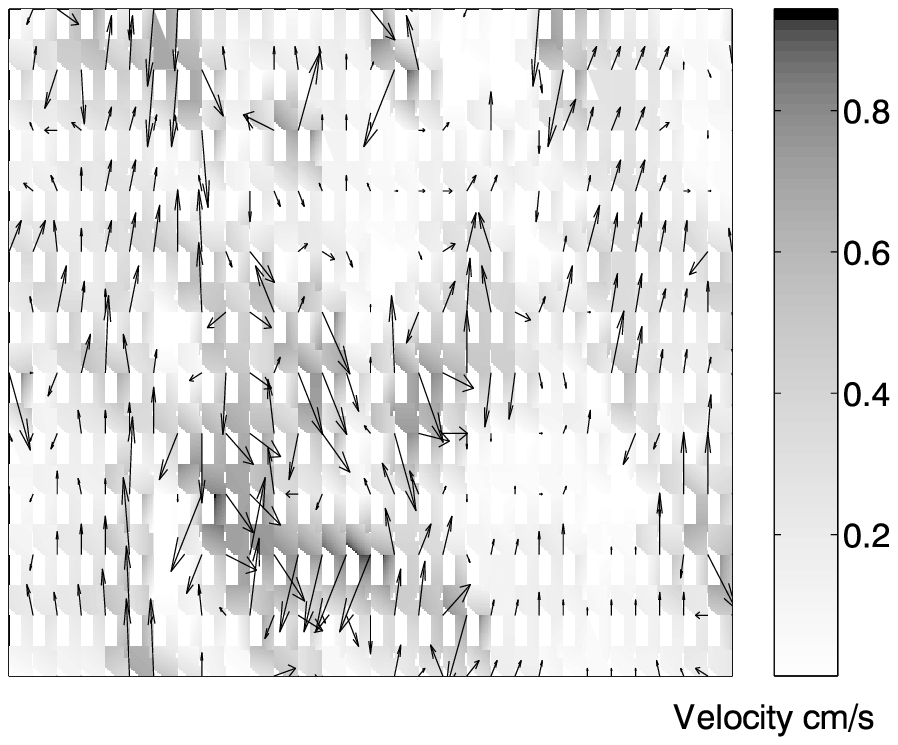} 
\caption{\small Multiple large scale flow cells at AR=0.65
  (a)Horizontal planform of plume structure at $Ra_{dc}=
  4.07\times10^{11},\,\frac{\Delta\rho}{\rho}_{dc}\,=\,4.52\times
  10^{-3}$. Image size is 15cms $\times$ 15 cms (b)Vertical plume
  structure at $Ra_{dc}=4.06\times
  10^{11},\,\frac{\Delta\rho}{\rho}_{dc}\,=4.5\times 10^{-3}$ image
  size is 15cm $\times$ 6.7 cms (c)Velocity distribution for(b) }
\label{fig:10jan}}\qquad\qquad
\parbox[t]{0.45\textwidth}{
\includegraphics[width=0.45\textwidth]{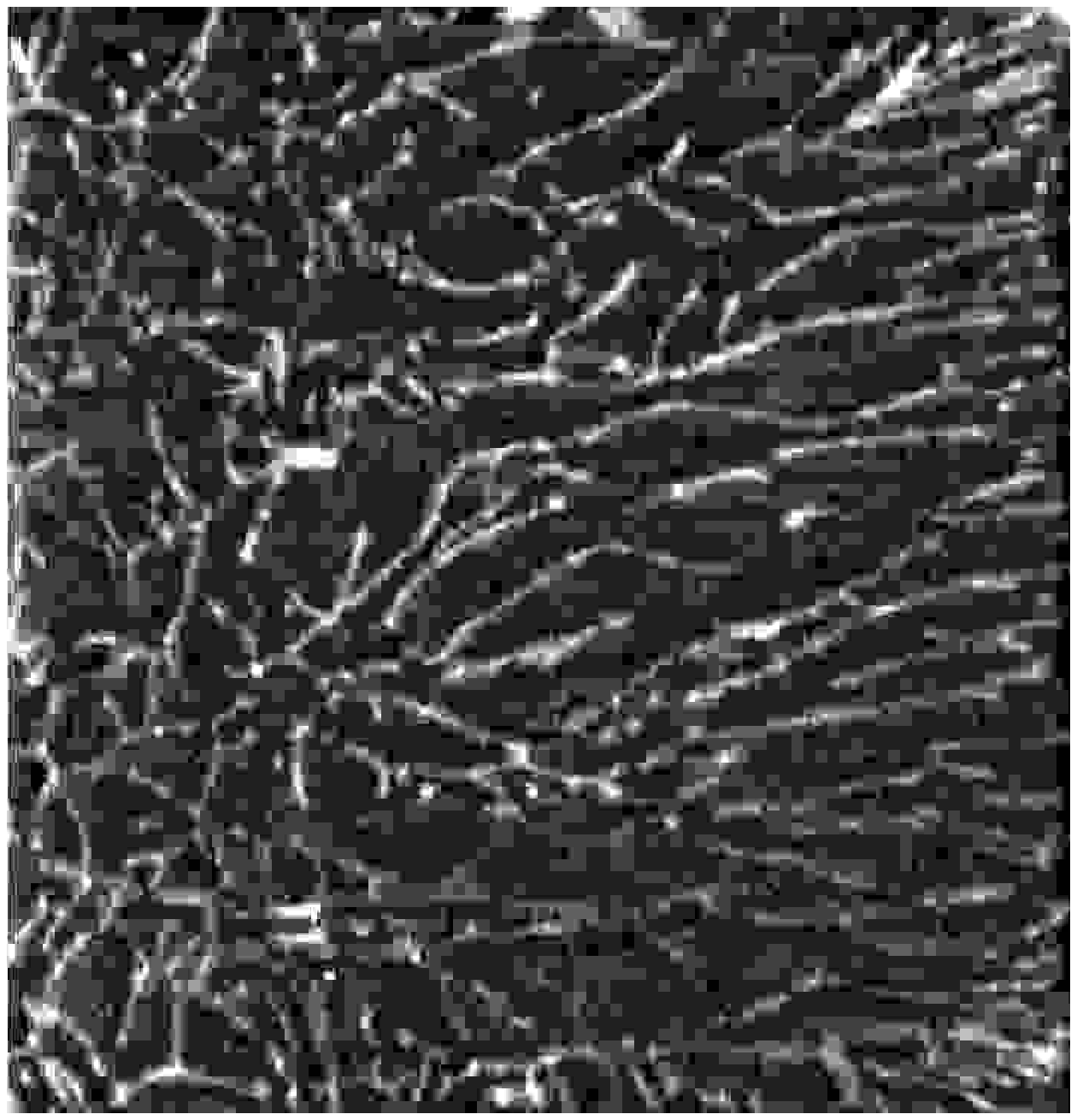} 
\includegraphics[width=0.3263\textwidth,height=0.2104\textwidth]{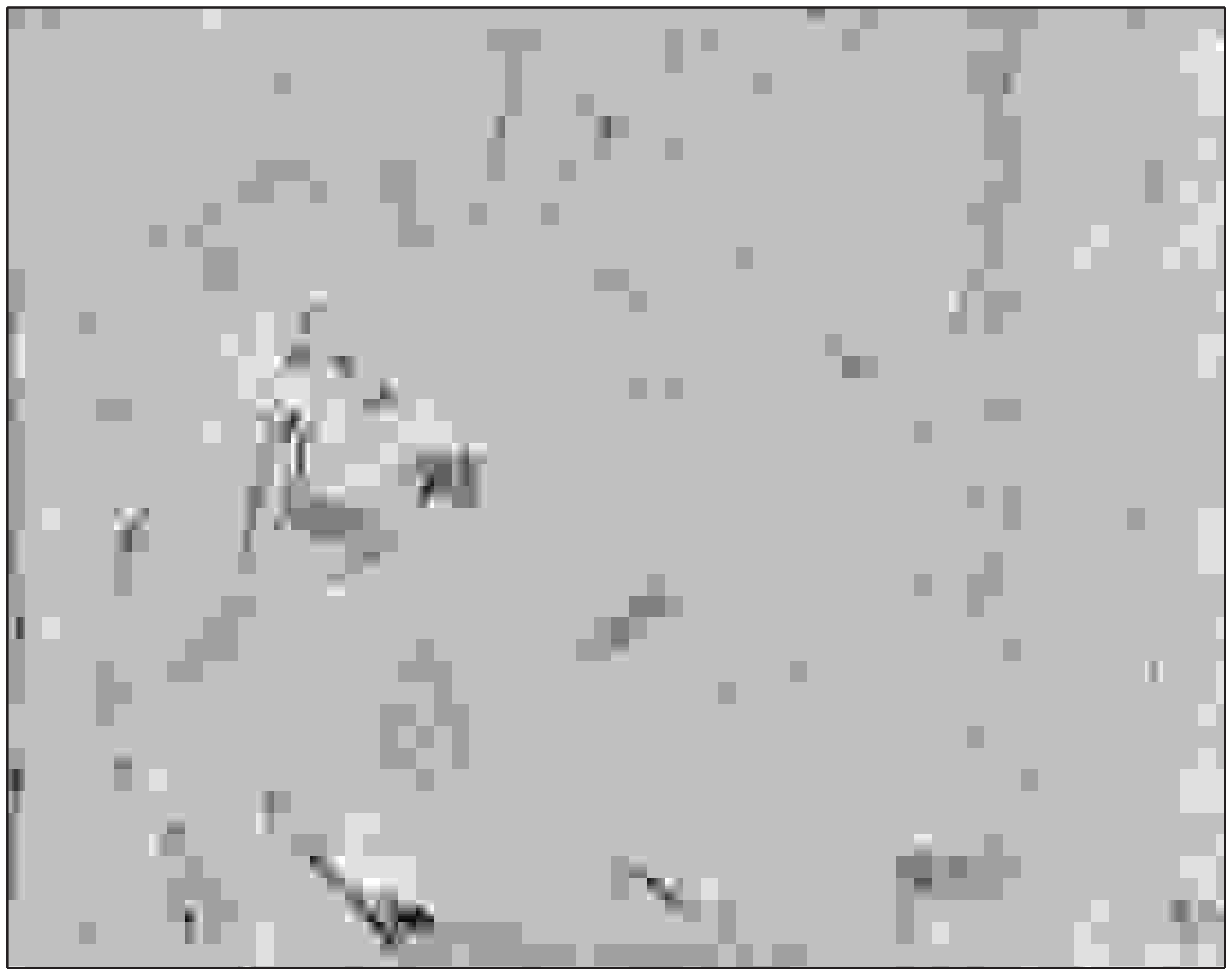}
\includegraphics[width=0.40\textwidth,height=0.2104\textwidth]{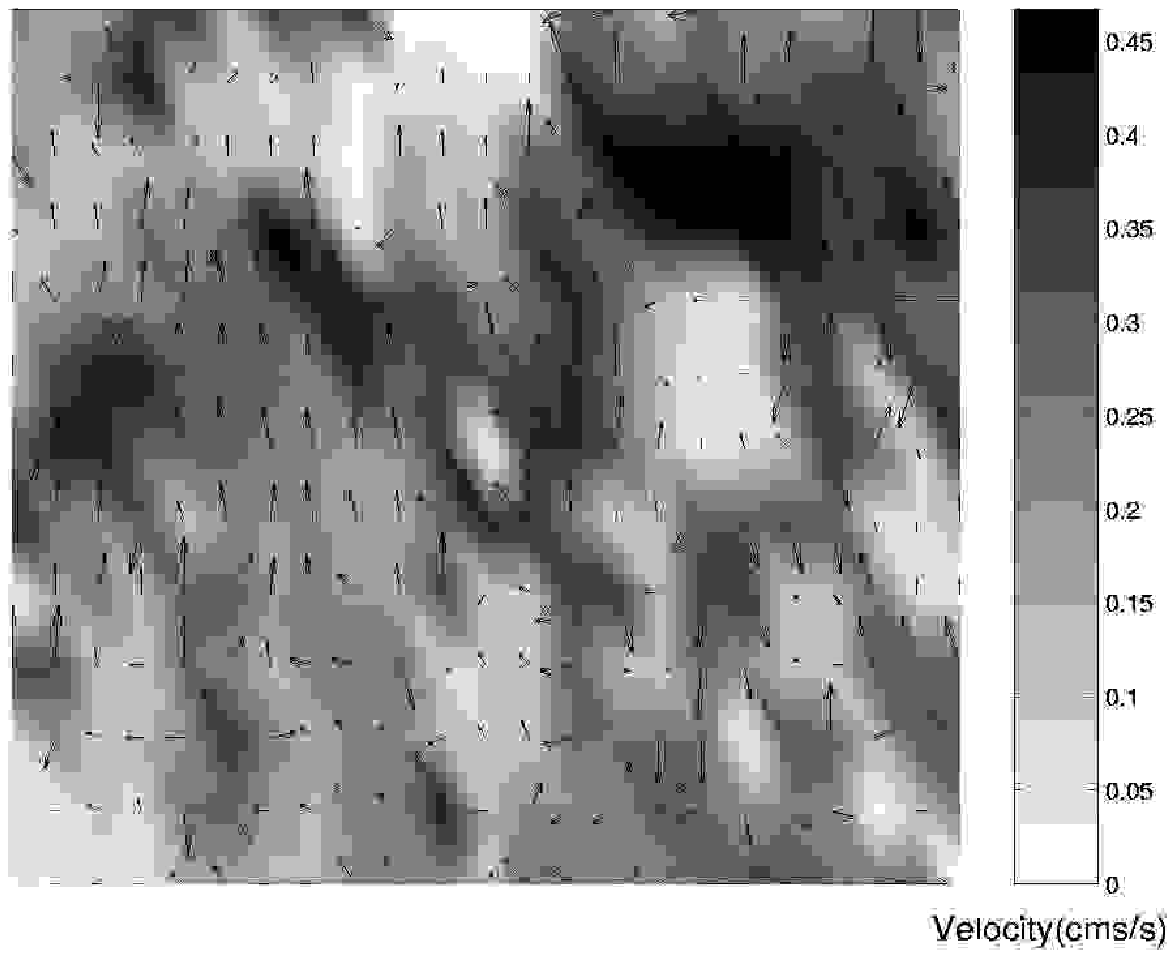} 
\caption{\small Single large scale flow cell at AR=0.435
  (a)Horizontal planform of plume structure at $Ra_{dc}=
  4.068\times10^{11},\,\frac{\Delta\rho}{\rho}_{dc}\,=\,4.515\times
  10^{-3}$ Image size is 10cms $\times$ 10 cms (b)Vertical plume
  structure at $Ra_{dc}=4.1003 \times
  10^{11},\,\frac{\Delta\rho}{\rho}_{dc}\,=4.551\times 10^{-3}$ Image
  size is 7.25cms $\times$ 4.67 cms (c)Velocity distribution for(b) }
\label{fig:18dec}}
\end{figure}
Figure~\ref{fig:10jan}(a) shows the plan form plume structure,
obtained when the upcoming plumes intersect a horizontal laser sheet
very near($<$1mm ) the membrane. The image shows the full test section
cross setion of 15cm $\times$ 15cm (AR = 0.65) and is at $Ra_{dc} =
4.0711\times 10^{11}$ and $\frac{\Delta\rho}{\rho}_{dc}\,=\,
4.6\times10^{-3}$ , where the flux follows $4/3^{rd}$ law (see figure
\ref{fig:radc}). The figure shows that the near wall coherent
structures in high Rayleigh number convection are line plumes.  We
notice that the plume structure display circular patches of aligned
lines originating from a plume free area in the center. The aligned
nature of the plume lines implies the presence of a near wall flow
along these lines. The predominant motion of the fluid in these
patches, as observed from the video images, were outward from the
plume free circular area along the plume lines. The lateral shift of
the plumes lines,due to the entrainment velocity of the nearby line
plumes, was also very low ($\sim $1mm/s).  Hence we infer these cells
as the signatures of large scale flow cells.  The large scale flow
impinges on the plume free circular area in the center of each aligned
plume region and create an outward near membrane flow which aligns the
line plumes.  The current image shows two large scale flow cells on
the right side of the image.  These large scale flow cells shift their
position randomly, implying that the large scale flow direction in
high Ra convection, at least for higher AR, is not constant.

The above picture would be more clear from Figure~\ref{fig:10jan}(b)
which shows the plume structure in a vertical plane at 2cms from a
side wall of the test section,( ie 2cms from the bottom of
figure~\ref{fig:10jan}(a)).  Note that Figure~\ref{fig:10jan}(a) is
after 11.5 minutes from figure~\ref{fig:10jan}(b), so that the latter
figure is not the exact vertical view of the former. The height of
figure~\ref{fig:10jan}(b) is 6.7 cms.  The figure shows that plumes
combine giving rise to columns of upward rising fluid, which results
in a downward travelling portion of fluid in between the columns,
which impinges on the wall and create the near wall shear.
Figure~\ref{fig:10jan}(c) shows the near wall velocity field
vectors(overlaid over the velocity magnitude) estimated using the
spatial intensity correlation technique\cite{cpg} between
figure~\ref{fig:10jan}(b) and another image 0.4 seconds later. The
upward moving colums and the resulting downward impingement in between
them could be noticed from the figure.  The column rise velocity is
about 0.3 cms/s while the downward velocites are much larger ($\sim$
0.8 cms/s). It is known that the large scale velocity scales as the
Deardorff's velocity scale \mbox{$W_*\,=\,\left(g\beta q h
  \right)^{1/3}$.  ~\cite{dear1}} $W_*$ for
Figure~\ref{fig:10jan}(c)is 0.31 cms/s, of the same order as the plume
rise velocity.  Hence, one could infer that the large scale flow
velocity is essentially driven by plume columns, which inturn organise
the plume structure. The current study shows that the mean shear near
the walls could be larger than that due to $W_*$ as the downward
velocities between the plume columns are higher than $W_*$ .

The planform plume structure at AR=0.435 is shown in
figure~\ref{fig:18dec}(a). The image size is the same as the test
section cross section of 10cms $\times$ 10 cms. Figure
\ref{fig:18dec}(b) shows the vertical plume structure 13 min prior to
figure~\ref{fig:18dec}(a). The image shows a vertical section(normal
to plume structure in figure~\ref{fig:18dec}(a)) 2cms above the bottom
side of figure~\ref{fig:18dec}(a) of 7.25cms width and 4.67 cms height
starting from the left of figure~\ref{fig:18dec}(a).  The plan form
plume structure and the vertical image lead us to the inference that
there is a single large scale flow cell spanning the full tank cross
section. The large scale flow impinges on the membrane on the right
and creates a near membrane velocity from right to left. This results
in inclined plumes near the membrane as seen in the vertical image.
Plumes combine and rise along the left wall, feeding the mean
circulation, which sustains the mean shear near the membrane. The
velocity estimates in figure~\ref{fig:18dec}(c) from spatial
correlation between images 0.4 seconds apart reproduces the dynamics
reasonably well, with the plume column rise velocity at the LHS of the
figure($\sim$ 0.3cms/s) being of the same order as the large scale
flow velocity estimated as the Deardorff scale (0.31cms/s). 
\section{Conclusions}
\label{sec:conclusions}
Experiments show that the flux scaling in high Rayleigh Number -high
Schmidt number turbulent convection, even in the presence of a large
scale flow, follows the 4/3$^{rd}$law and seems to have little
dependence on Sc. The flux is only weakly dependent on the AR. The
near wall coherent structures are line plumes. We detect multiple
large scale flow cells and changing large scale flow direction at
higher AR. Lower AR shows a single large scale flow cell and constant
sense of near membrane mean shear. The large scale flow is shown to be
sustained by rising columns of combined plumes, the velocity of which
scales as Deardorff scale.

\bibliography{babu} \small
\end{document}